\makeatletter
\newcommand{\newoperator}[2]{\@ifdefinable#1{\def
#1{\mathop{#2}\nolimits}}}
\makeatother

\def\gth{\theta}

\newoperator{\dif}{{\rm d\!}}
\newcommand{\deriv}[3]{\frac{{\rm d}^{#1}{#2}}{{\rm d}{#3}^{#1}}}

\documentclass{aa}

\usepackage{graphicx}
\usepackage{dcolumn}
\usepackage{amssymb}

\newcolumntype{d}[1]{D{.}{.}{#1}}

\begin{document}

\title{Testing general relativity by micro-arcsecond global astrometry}

\author{A.~Vecchiato \inst{1}${}^,$\inst{3}${}^,$\inst{4} \and M. G.~Lattanzi \inst{2} \and
B.~Bucciarelli \inst{2} \and M.~Crosta \inst{2}${}^,$\inst{3}${}^,$\inst{4} \and
F.~de~Felice \inst{1}${}^,$\inst{3}${}^,$\inst{4} \and M.~Gai \inst{2} }

\offprints{lattanzi@to.astro.it}

\institute{Istituto di Fisica ``G. Galilei'', Via Marzolo 8, I-35100 Padova, Italy
      \and INAF, Osservatorio Astronomico di Torino, I-10025 Pino Torinese TO, Italy
      \and Istituto Nazionale di Fisica Nucleare (INFN), Sezione di Padova
      \and Centro per gli Studi e le Attivit\`a Spaziali ``G. Colombo'' (CISAS), Padova}

\date{Received ; accepted }

\abstract{The global astrometric observations of a GAIA-like
satellite were modeled within the PPN formulation of
Post-Newtonian gravitation. An extensive experimental campaign
based on realistic end-to-end simulations was conducted to
establish the sensitivity of global astrometry to the PPN
parameter $\gamma$, which measures the amount of space curvature
produced by unit rest mass. The results show that, with just a few
thousands of relatively bright, photometrically stable, and
astrometrically well behaved single stars, among the $\sim10^9$
objects that will be observed by GAIA, $\gamma$ can be estimated
after 1~year of continuous observations with an accuracy of
$\sim10^{-5}$ at the $3\sigma$ level. Extrapolation to the full 5-year mission
of these results based on the scaling properties of the adjustment procedure
utilized suggests that the accuracy of $\simeq 2\cdot 10^{-7}$, at the same $ 3\sigma $ level,
can be reached with $\sim 10^6$ single stars, again chosen as the most astrometrically stable
among the millions available in the magnitude range $V=12-13$.
These accuracies compare quite favorably with recent findings of scalar-tensor
cosmological models, which predict for $ \gamma$ a present-time
deviation, $\left| 1 - \gamma \right|$, from the General
Relativity value between $ 10^{-5} $ and $ 10^{-7} $.

}

\titlerunning{Astrometric test of General Relativity}
\authorrunning{Vecchiato et al.}

\maketitle


\section{Introduction}
Since General Relativity's (GR) first appearance, many alternative
gravity theories have been proposed. Experiments were then needed not
only to test the validity of GR against Newton's theory, but also every
alternative theory of gravity against all the others (Will \cite{will93},
\cite{will01}).

The Parametrized Post-Newtonian formalism (PPN) takes the slow-motion,
post-Newtonian limit of all the metric theories and exploits their
similarities to give them a unique, coherent framework by introducing a
set of 10 PPN parameters. Each theory is then characterized by a particular
value for each of the PPN parameters.

The validity of the post-Newtonian expansion is confined to those
physical situations in which the energy density, regardless to its
specific form, is small. Considering the various physical cases,
this means that gravity fields ($U$), velocities ($v$), pressure
to matter density ratios ($P/\rho$), and internal specific energy
densities \( \Pi \), are all small quantities so that (Ciufolini
\& Wheeler \cite{ciufwheel}; Will \cite{will93})
\[
\epsilon^2\equiv\frac{v^{2}}{c^{2}}\sim \frac{U}{c^{2}}\sim
\frac{P}{\rho c^{2}}\sim \Pi \ll 1,
\]
where $\epsilon \ll 1$ is the ``smallness parameter''. The
slow-motion hypothesis also implies that, for any quantity $A$,
its time derivative is much smaller than its space derivative, so
that (Misner et al. \cite{misnetal})
\[
\frac{\left| \partial A/\partial t\right| }{\left| \partial
A/\partial x\right| }\sim \epsilon ;
\]
this makes the PPN framework particularly convenient for Solar
System experiments (where e.g. \( U/c^{2}\lesssim 10^{-5} \)).
Each experiment allows one to measure the values of some PPN
parameters, so that the theories that do not match those values
are ruled out.

The most investigated among the PPN parameters are $\gamma$ and
$\beta$, which measure respectively the amount of space-curvature
produced by unit rest mass and the amount of non-linearity in the
superposition law of gravity.

In GR the parameters $\gamma$ and $\beta$ are both
equal to 1, while other theories aiming at the formulation of a
unified theory predict small deviations from the GR value.  The
most promising of such theories consider the existence of a scalar
field which, along with the usual metric tensor, participates in
the generation of gravity. For this reason, such theories are
usually called {\em scalar-tensor\/} theories.  As a consequence,
the value of $ \gamma $ predicted by the scalar-tensor theories of
gravity deviates from 1, and it is generally an
adjustable value.  Moreover, Damour and Nordtvedt
(\cite{damnordt}) showed that scalar-tensor cosmological models
contain a sort of attractor mechanism toward GR.
In practice the value of $\gamma $ changes with time and, whatever
the value at the birth of the universe, it tends to 1 as $
t\to\infty $.  The present-time value of $ |1-\gamma| $ (i.e. its
deviation from GR) depends on the efficiency of the attractor
mechanism, and Damour and Nordtvedt calculate a value between $
10^{-5} $ and $ 10^{-7} $.  More recently, Damour et al.~(\cite{dametal02a},
\cite{dametal02b}) have given a new estimation of $ |1-\gamma| $,
within the framework compatible with string theory and modern
cosmology, which basically confirms the previous results$^($\footnote{
Within this framework $ |1-\gamma|\simeq-2\alpha_\mathrm{had}^2 $, where
$ \alpha_\mathrm{had} $ is the dilaton coupling to hadronic matter.
Its value depends on the model taken for the inflation potential
$ V(\chi)\propto\chi^n $, $ \chi $ being the inflation. So the range
of the expected deviations from GR is between $ \sim3.6\cdot10^{-5} $
($ n=4 $) and $ \sim0.5\cdot10^{-7} $ ($ n=2 $) (Damour et al.~\cite{dametal02b}).
}$^)$.

Although optical astrometry provided the first direct test of
GR by measuring the amount of light bending due to
the gravitational pull of the Sun (Dyson, Eddington, and Davidson
1920), observing conditions from the ground have severely limited
the possibility of this technique to contribute to the search for
a scalar component of gravity. Radio astrometry has been able to
achieve significantly better results, however it is again the
observing conditions (especially the effects of ionosphere and
troposphere) which are ultimately limiting its accuracy in
measuring gravitational light deflection from the ground. Presently, $\gamma$
is known to an accuracy of $\sim 10^{-3}-10^{-4}$ (Will \cite{will01}), therefore
well above the level necessary to detect the scalar component
predicted by Damour and Nordtvedt.

The success of the mission Hipparcos has proven that going into
space is the way for increasing the accuracy of astrometric
measurements to more interesting values. By utilizing the
measurements of Hipparcos, (Froeschl\'{e} et al. \cite{froeschle}; see also Mignard
~\cite{mignard}) were able to derive $\gamma$ with an accuracy comparable to
that of the best radio experiment (Lebach et al. ~\cite{lebach}). Therefore,
expectations are that the micro-arcsec ($\mu$-arcsec) accuracy
targeted by the space missions SIM, approved by NASA, and GAIA,
approved by ESA, will be able to measure $\gamma$ with a precision
of 10$^{-5}$--10$^{-6}$ (Danner and Unwin \cite{danunw}) or
better (ESA \cite{redbook}, de Felice et al. \cite{defelice3}),
positioning optical astrometry at the fore front of experimental
gravitation.

In this article we focus our attention on the GAIA project, a
space astrometry mission that has recently been confirmed as a
Cornerstone mission of the ESA program of scientific satellites.
GAIA is expected to be launched not later than 2012, with a
possible window of opportunity in 2010. For this reason ESA has
devised a mission implementation plan that will be able to cope
with a 2010 launch.

GAIA is a scanning astrometric satellite which builds on the
successful concept of the Hipparcos mission, but with
order-of-magnitude improvements in, e.g., the number of objects
observed and measurement precision. It will be essential for the
scientific advancement in many branches of astronomy, and
especially in the fields of stellar astrophysics and galactic
astronomy (Perryman et al. \cite{macp}). Here we report on the
first thorough attempt at determining the accuracy with which GAIA
could measure the parameter $\gamma$. This is done through a
realistic end-to-end simulation which considers the satellite most
relevant modes of observation, the relativistic environment in
which such observations take place, and the expected single-measurement errors.

\section{The Parametrized Post-Newtonian (PPN) Model}
Using the Eddington-Robertson form of the PPN formalism (where only the parameters
$\gamma$ e $\beta$ are considered),
we have developed a model based on the assumptions that the only
source of gravity is a spherical and non-rotating Sun and the
observer is moving on a circular orbit. This scenario is
consistent with that discussed in our previous papers (de Felice
et al. \cite{defetal1}, \cite{defetal2}; hereafter Paper I and
Paper II, respectively),where a non-perturbative approach was
developed as a first attempt at modeling the GAIA observations in
a rigorous relativistic environment. Therefore, the results discussed in
this paper can directly be compared with the findings reported in
those earlier works.

Under the above hypotheses the PPN metric becomes (Misner et al.,
\cite{misnetal})
\begin{eqnarray}
\label{eqn:PPN-metric}
\mathrm{d}s^{2}&=& -\left[ 1\! -\! 2\frac{M}{r}\! +\! 2\beta \left( \frac{M}{r}\right) ^{2}\right] \mathrm{d}t^{2}+
                   \nonumber\\[7pt]
               &{}&\left[ 1\! +\! 2\gamma \frac{M}{r}\right]\left[ \mathrm{d}r^{2}\! +\! r^{2}\left( \mathrm{d}\theta ^{2}\! +\! \sin ^{2}\theta \, \mathrm{d}\phi ^{2}\right) \right]
\end{eqnarray}
where $r$, $\theta$, and $\phi$ are spherical coordinates centered
on the Sun, $t$ is coordinate time, and $M$ is the geometrized mass
of the Sun.

As observable we consider the cosine of the angle $\psi_{12}$
between two stars; it can be expressed as
\begin{equation}
\label{eqn:scal-rel}
\cos \psi _{12}=\frac{h_{\alpha \beta }k_{1}^{\alpha }k_{2}^{\beta }}{\sqrt{h_{\iota \pi }k_{1}^{\iota }k_{1}^{\pi }}\sqrt{h_{\rho \sigma }k_{2}^{\rho }k_{2}^{\sigma }}},
\end{equation}
where \( k_{1}^{\alpha } \) and \( k_{2}^{\alpha } \) are the
four-velocities of the photons reaching the observer (i.e. the
four-vectors tangent to the photons' null-geodesics) and
\( h_{\alpha \beta }=g_{\alpha \beta }+u_{\alpha }u_{\beta } \) is a
tensor which projects to the rest frame of the observer, i.e., the GAIA satellite. Here
\( g_{\alpha \beta } \) and \( u_{\alpha } \) are the space-time metric and the observer's
four-velocity, respectively.


As expected, the expression we found for the four-velocities of
the photons $k^{\alpha}$ depends on both $\gamma$ and $\beta$;
then, they could be made part of the data reduction process as
unknown parameters, in addition to those describing position and
velocity of the stars. However, $\beta$ enters the $k^{\alpha}$
as a second order term compared to $\gamma$ (see also
Eq.(\ref{eqn:PPN-metric})); therefore, it was decided not to
consider $\beta$ in the derivation of the observation equations
(see below) but to set it to 1, namely to its value in GR.

Following the method developed in Paper I and II, and after a lengthy
calculation, we then obtained the linearized observation equation as
function of the astrometric parameters (i.e., angular coordinates,
parallaxes, and proper motions), and of $\gamma$ in the form (Vecchiato
\cite{vecchiato})
\begin{eqnarray}
   -\sin\psi_{12}(t)&\delta\psi_{12}&=
          A_1\,\delta\gth_1(t_0)+B_1\,\delta\phi_1(t_0)+
            \nonumber\\[7pt]
      &{}&C_1\,\delta p_1+D_1\,\delta\mu_{\gth_1}+E_1\,\delta\mu_{\phi_1}+
   \nonumber \\[7pt]
      &{}&A_2\,\delta\gth_2(t_0)+B_2\,\delta\phi_2(t_0)+
            \nonumber\\[7pt]
      &{}&C_2\,\delta p_2+D_2\,\delta\mu_{\gth_2}+E_2\,\delta\mu_{\phi_2}+
            \nonumber\\[7pt]
      &{}&F\,\delta\gamma,
   \label{eqn:cond.eq.}
\end{eqnarray}
where the coefficients are derived from the differentiation of the
right-hand side of Eq.(\ref{eqn:scal-rel}) with respect to each of the
unknowns.

As in Paper II, parallax and (angular) proper motions are defined as
\[
p\equiv r_\oplus/r, \qquad \mu_\gth \equiv \deriv{}{\gth}{t}, \qquad \mu_\phi \equiv \deriv{}{\phi}{t};
\]
here $r_\oplus$ is the Earth's mean orbital radius, $r$ is the
coordinate radial distance of the star from the Sun and $t$ is the
coordinate time.

\section{The end-to-end simulation}
The simulation follows the procedure used in the previous
non-perturbative works (Paper I and II), the main change being the
presence of the new unknown $\gamma$ that modifies the design
matrix of the system of condition equations.

First, we generate the set of {\it true} quantities, which define initial location and temporal
evolution of the stellar positions on the celestial sphere and the true value of $\gamma$. This
was set to its GR value, i.e., $\gamma$ = 1. The corresponding {\it catalog} values
are calculated, as usual, by adding suitable root-mean-square (rms) errors to the true values. Next,
we find the stellar pairs which can be observed by a satellite that sweeps the sky following a
Hipparcos-like scanning law. Once the stellar pairs are known the true angular distances (from the
true coordinates) are calculated, and the satellite observations are generated by perturbing these
true arcs with the observational error in Table~\ref {tab:input-param}. The catalog arcs (from the
catalog coordinates) are also computed at this stage. Only arcs joining stars lying in different
fields of view (FOVs) were counted without degeneration.

The result of these two steps is the generation of the measured quantity, $-\sin\psi(t)\delta\psi$,
and of the coefficients in Eq.~(\ref{eqn:cond.eq.}) for the construction of the linearized condition
equations for all of the pairs observed during the mission lifetime we decided to simulate. Finally,
the least-squares solution of the system is found by means of a conjugate-gradient method, suitable
for large and sparse matrices like ours, and the errors are computed by direct comparison
to the true values.

The least-squares solution returns the estimates of the adjustments to the catalog values of the
unknowns in Eq.~(\ref{eqn:cond.eq.}), i.e.,
$\tilde{\delta\theta}$, $\tilde{\delta\phi}$, $\tilde{\delta\mu_\theta}$, $\tilde{\delta\mu_\phi}$,
$\tilde{\delta p}$, and $\tilde{\delta\gamma}$. For the
adjustment to the $\gamma$ parameter, the corresponding rms error is indicated with the symbol
$\sigma_{\delta\gamma}$. Notice that this error is the same as the error of $\gamma$,
$\sigma_\gamma$,
as $\tilde{\gamma} = \gamma_{cat} + \tilde{\delta\gamma}$, where $\gamma_{cat}$ is the simulated catalog
value of the deflection parameter. The symbols $\sigma_{\delta\gamma}$ and $\sigma_{\gamma}$ are both
used in the reminder of this article.

\section{Experiments and results}
\begin{table*}
\begin{tabular}{ccc}
   \hline\vspace{5pt}
   {\em Parameter} & {\em Numerical value} & {\em Comment} \\
   \hline
   orbital radius & $1.496\cdot10^{11}$~m & same as Earth's orbital radius
                                            ($R_\oplus$) \\[3pt]
   precession angle & $43^\circ$ & same as solar aspect angle \\[3pt]
   precession speed of the spin axis & 6.4~rev/yr & \\
   satellite spin period & 128~min & \\
   angle between the & $54^\circ$ & \\
   viewing directions & & \\
   field-of-view & $1^\circ\!\!.6$ & \\
   of each telescope & & \\
   mission starting time ($t_0$) & $-T/2$ & minimum correlation between \\
                                 &        & coords. and proper motions \\
   radius of the simulated sphere & 2~mas & uniform density sphere \\
                                  &       & of 500~pc in radius \\
   catalog error on & 2~mas & \\
   coordinates and parallax & & \\
   catalog error on $\gamma$ & $\sigma_\gamma=2\cdot10^{-3}$ & \\
   single-measurement error ($\sigma_{obs}$) & 10~$\mu$arcsec & as expected
                                                                for pairs of \\
                                              & & V $\sim 12-13$ ~mag stars \\
   \hline
\end{tabular}
\caption{ Most relevant parameters common to all of the simulations. $T$ is the mission duration. For the
dynamical simulations, the stars are generated within a uniform density sphere of 500~pc centered on
the Sun (i.e., $p\geq 2$~mas). Of course, the distribution of the resulting sample of simulated
stars is representative only of a (relatively) small portion of the actual Galaxy. Nevertheless, as
mentioned in Paper II, the adopted values are sufficiently realistic for the immediate scope of this
work, which is to gauge the sensitivity of GAIA to the estimation of $\gamma$. }
\label{tab:input-param}
\end{table*}

\subsection{An upper limit on the accuracy of $\gamma$}
Before going into the details of our experimental campaign it is
useful to show, through a simple order-of-magnitude calculation,
what kind of accuracy can be expected for $\gamma$ with GAIA-like
observations. This calculation starts from the consideration that
the satellite measurements for the estimation of $\gamma$ can be
thought of as ``Eddington-like'' measurements of very high precision
but with the stars at some tens of degrees from the solar limb.
Each observation contributes to the determination of $\gamma$ with a
precision of $\Delta_\gamma\sim10^{-2}-10^{-3}$~${}^($\footnote{ This
can be seen by taking the expression for the light deflection of a
light source (e.g. Misner et al., \cite{misnetal}), solving for
$\gamma$ and then applying the error propagation formula. Then, a
single observation of a star at $45^\circ$ from the Sun with
$\sigma_{obs}=10~\mu$arcsec, yields
$\Delta_\gamma\simeq2\cdot10^{-3}$. }${}^)$. If $N$ is the number of
such observations, the final accuracy will be approximately
$\sigma_\gamma\sim\Delta_\gamma\cdot N^{-1/2}$. For a period of,
say, 1.5 years of continuous operations and 6500 of the stars in
the magnitude range V=12-13 mag, GAIA  would provide (see below)
about 450000 observations, thus $\sigma_\gamma\sim3\cdot10^{-6}$.
In the calculation above we have disregarded the facts that the geometry
of the GAIA observations is different (the gravitational deflection
is ``seen" through its differential effect along the arcs joining the star
pairs) and that $\gamma$ is not the only unknown to be estimated. Therefore,
that value of $\sigma_\gamma$ is clearly representative of a best case scenario, and
it is used only for comparison in the following discussion.

The simulation campaign was split in two parts. The first part was
devoted to test the new code written for the PPN model and to make
sure that the results were compatible with those in Paper I and
II, and with the empirical prediction for $\gamma$ derived above. The second set
of experiments was destined to establish the relation describing
the GAIA sensitivity to $\gamma$  as function of the number of
stars. Such a relation can then be used to make predictions on the
accuracy of $\gamma$ for any given number of stars.

\subsection{Validation of the PPN model}
Table \ref{tab:input-param} lists the values of the input parameters
which were common to all of the simulation runs. The main difference with our
earlier work is in the single-observation error $\sigma_{\rm obs}$, which
is set here to 10~$\mu$arcsec. This is the value expected for the error of
one arc joining pairs of equal-magnitude stars, approximately 5 magnitudes
brighter than those utilized in our previous experiments, i.e., V$\sim$12 mag.
The 10-$\mu$arcsec error is compatible with current estimates of the GAIA error
budget for stars brighter than $V=12-13$ (ESA, \cite{redbook})${}^($\footnote{As mentioned
in Paper I, it is assumed that the physical properties of the stars considered in these
experiments are such that they do not show any intrinsic astrometric noise which
adds to the measurement error. For example, single and non-variable solar-type
dwarfs would be ideal targets.}${}^)$.

As for the $\gamma$ parameter, the starting (catalog) value for the PPN
parameter is generated from the true value by adding an error of $\sigma_\gamma=2\cdot10^{-3}$,
which is comparable with current best estimates.

The new code was tested on different sets of 50 simulations,
and each Monte-Carlo set was run with the same values of the input parameters to have
statistically significant results. The results of this series of experiments is
summarized in Table \ref{tab:res}.

The first row is representative of the runs with a ``static sphere", i.e., the ideal situation where all the
stars have no intrinsic motions and are located at such a large (infinite) distance that the
parallactic motion due to the observer is also null. In this case, the location on the celestial sphere
of the simulated stars is completely specified by their angular coordinates ($\theta$ and $\phi$).
Therefore, each linearized observation equation, Eq.\ref{eqn:cond.eq.}, has  only five unknowns,
the four corrections to the  angular coordinates of each star pair, and the adjustment to the $\gamma$
parameter.

The second row presents the results of the simulations with a ``dynamical sphere", i.e., the case
where the stars move with time. For simplicity, only the parallax motion was simulated; the stars
were generated within a uniform density sphere of 500~pc centered on the Sun, i.e., with parallaxes
$p\geq2$~mas (200 times the measurement error), and, as for the static case, with no intrinsic
(cosmic) motion. Notice that the mission duration was increased to 2~years. According to the findings
in Paper II, this is the minimum observation period required for a reliable reconstruction of the
dynamical parameters.

\begin{table}
\[
\begin{array}{c|c|c|c|c|c|c|c|c}
T   & n^*  & n_{obs} & Q     & \sigma_{obs}/\sqrt{Q} & \sigma_{\delta p} & \sigma_{\delta\theta} & \sigma_{\sin\theta\,\delta\phi} & \sigma_{\delta\gamma}\cdot10^5 \\
\hline
1.5 & 6500 & 453071  & 34.85 & 1.69                  & --                & 1.76                  & 1.88                            & 0.94                  \\
2.0 & 5000 & 356061  & 23.74 & 2.05                  & 2.05              & 1.31                  & 1.75                            & 1.92                  \\
\end{array}
\]
\caption{Summary of the results of the different sets of simulations utilized to test the new code developed
for the PPN model. $Q$ is the number-of-observations to the number-of-unknowns ratio.}
\label{tab:res}
\end{table}

As expected, the true errors of the astrometric parameters given in Table~\ref{tab:res} compare quite
well with what was obtained in Papers I and II after taking into account the differences in the values
of $Q$ and the factor of 10 in measurement errors, consistent with the 5-mag difference of the stars
considered in the new experiments. In particular, the results reported in the last row of Table 2 of Paper II,
reproduce very closely what shown in our Table~\ref{tab:res} for a mission duration of $T$=2 years.

\subsection{What accuracy on $\gamma$?}
Of much greater interest is the fact that the errors of $\gamma$ after the sphere reconstructions shown
in Table~\ref{tab:res} are considerably close to the best-case value of $\sim$ 3$\cdot$~10$^{-6}$ derived
at the onset of this section (see also Table 4 in de Felice et al 2000).

This encouraging result brought us to consider a new set of simulations with the intent to study the
accuracy of $\gamma$ for larger samples of stars. Indeed, the all-sky survey nature of the GAIA
observations ensures that all of the objects in the magnitude range of interest will be observed during
the operational life of the satellite. This means that, although the actual number will considerably reduce
because of the stringent requirements on the intrinsic astrometric stability of the sources, millions of
potential targets will be available for the ``$\gamma$ experiment".
Unfortunately, the computing power needed to perform a data reduction simulation of such a size is beyond
our present resources, therefore we had to resort to an alternative schema to find the desired answer.
We generated 11 simulations of a static sphere, each consisting, as before, of 50 runs with the same initial
conditions and the mission duration set to 1~yr. The number of stars was increased from $n^*$=5000 to the
maximum extent possible, i.e., $n^*$=15000. The results are listed in Table~\ref{tab:res-gamma}.

\begin{table}
\begin{tabular}{|c|c|c|}
\hline
\( n^{*} \)&\( <n_{\mathrm{obs}}> \)&\( \sigma_{\delta\gamma} \)\\
\hline
\hline
 5000&\( 1.78\cdot 10^{5} \)&\( 1.99\cdot 10^{-5} \)\\
\hline
 6000&\( 2.56\cdot 10^{5} \)&\( 1.91\cdot 10^{-5} \)\\
\hline
 7000&\( 3.49\cdot 10^{5} \)&\( 1.23\cdot 10^{-5} \)\\
\hline
 8000&\( 4.56\cdot 10^{5} \)&\( 1.33\cdot 10^{-5} \)\\
\hline
 9000&\( 5.79\cdot 10^{5} \)&\( 1.34\cdot 10^{-5} \)\\
\hline
10000&\( 7.14\cdot 10^{5} \)&\( 9.79\cdot 10^{-6} \)\\
\hline
11000&\( 8.63\cdot 10^{5} \)&\( 9.19\cdot 10^{-6} \)\\
\hline
12000&\( 1.03\cdot 10^{6} \)&\( 1.07\cdot 10^{-5} \)\\
\hline
13000&\( 1.20\cdot 10^{6} \)&\( 8.16\cdot 10^{-6} \)\\
\hline
14000&\( 1.40\cdot 10^{6} \)&\( 7.87\cdot 10^{-6} \)\\
\hline
15000&\( 1.60\cdot 10^{6} \)&\( 6.19\cdot 10^{-6} \)\\
\hline
\end{tabular}
\caption{Computed errors on the estimation of the PPN parameter $ \gamma $.
The numbers refer to the results of the eleven $ T=1 $~yr Monte-Carlo simulations with
increasing $ n^* $ (number of stars). In the second column $ <n_{\mathrm{obs}}> $
is the mean number of observations for the 50 simulations.}
\label{tab:res-gamma}
\end{table}

A single simulation, of the 50 comprising each Monte-Carlo set, yields one value of the difference
\( \delta \gamma =\gamma^{*}-\tilde{\gamma} \)(true \( \gamma \) minus estimated \( \gamma \));
each simulation is then a measure of $\gamma^*$ affected only by random errors, and the sample of
50 $\delta \gamma$'s has a Gaussian distribution. The standard deviation of this distribution,
$\sigma_\gamma$, is the measure of the error on $\gamma$ of each Monte-Carlo group of
simulations (Table~\ref{tab:res-gamma}). For example, if a difference of $3\cdot10^{-5}$ from
the GR value, $\gamma^*$=1, is measured for $\gamma$ with an error of
$\sigma_\gamma\simeq6\cdot 10^{-6}$, this would be interpreted as a 5-$\sigma$ detection of a
deviation from General Relativity.

\begin{figure}
\resizebox*{0.9\columnwidth}{!}{\rotatebox{-90}{\includegraphics{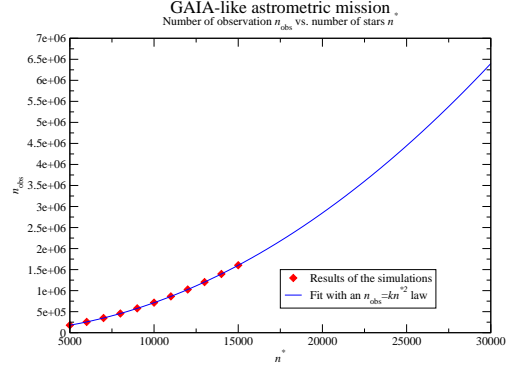}}}
\caption{Number of observations versus number of stars. The solid curve represents the
interpolated law $n_{\mathrm{obs}}$=$k \cdot {n^*}^2$, $k$=0.0071264.}
\label{fig:nobs-vs-nstar}
\end{figure}

\begin{figure}
\resizebox*{0.9\columnwidth}{!}{\rotatebox{-90}{\includegraphics{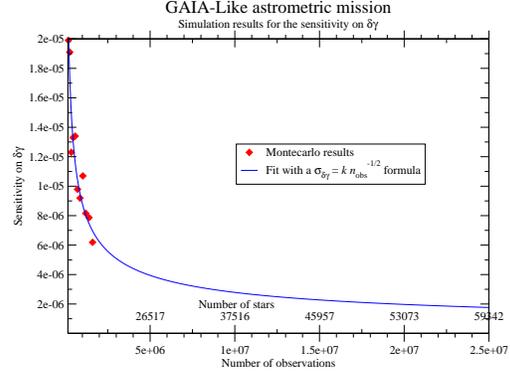}}}
\caption{$\sigma_{\delta\gamma}$ versus number of observations. The solid curve represents
the fit to the relation $\sigma_{\delta\gamma}=k\cdot n_{\mathrm{obs}}^{-1/2}$,
$k=0.0088179\pm 0.0000097$.}
\label{fig:sdg-vs-nobs-nstar}
\end{figure}

Fig.~\ref{fig:nobs-vs-nstar} shows the relation $n_{\mathrm{obs}}$= $k \cdot {n^*}^2$ we derived from
interpolating the data in Table~\ref{tab:res-gamma}. Also, simple statistical considerations suggest
to fit the results in columns 2 and 3 of Table~\ref{tab:res-gamma} to the
$\sigma_{\delta\gamma}\propto n_{\mathrm{obs}}^{-1/2}$ relation, and the result is shown in
Fig.~\ref{fig:sdg-vs-nobs-nstar}.

We are now in a position to make realistic predictions on the
error of $\gamma$ for a much larger number of stars
(observations), comparable to the size of the stellar
sample expected to be surveyed by the satellite at the magnitude
limit of interest here. For $\sim10^6$ stars ($\simeq$ $9.8\cdot
10^8$ observations), the value of $\sigma_{\delta\gamma}$
estimated from the curve in Fig.~\ref{fig:sdg-vs-nobs-nstar} is
$\simeq 9\cdot 10^{-8}$. This means that a 1~yr-long GAIA-like mission could measure a
value of $3 \cdot 10^{-7}$ for $\left|1 - \gamma \right|$, which would then represent
a $3 \sigma$ detection of the deviation of the actual gravitation from GR.
And at the end of the satellite lifetime, after 5 years of continuous observations,
the sensitivity to a $3\sigma$ detection will further improve by a
factor of $1/\sqrt{2}$ to $\sim 2 \cdot 10^{-7}$.
These numbers imply that GAIA could reach the lower bound of the interval of possible deviations
predicted by Damour and Nordtvedt (\cite{damnordt}) and by Damour et al.
(\cite{dametal02b}).

\subsection{How do we compare with the Hipparcos result? }
A Monte-Carlo experiment analogous to those above but with the relevant mission parameters set to
the values of the Hipparcos mission was also conducted. We deemed it as quite important to compare our
findings to the work of Froeschl\'{e} et al. (1997), who attempted the first direct determination
of $\gamma$ by means of the global astrometric data taken with the Hipparcos satellite.
Indeed, this would add confidence on the ability of our simulation to make realistic predictions on
the possibility to derive a very accurate value for $\gamma$ with GAIA.

As ``reference" experiment we adopted the case with 44000 stars in Table 1 of Froeschl\'{e} et al.
(\cite{froeschle}), which resulted in an error on the deflection parameter of
$\sigma_\gamma=4\cdot 10^{-3}$.

We first simulated $n^*$=15000 stars and an observing period of 1 year; the observational
and catalog errors were set to the values used in the Hipparcos experiments, i.e.,
$\sigma _{\mathrm{obs}}=3\, \mathrm{milli-arcsec}$ and $\sigma _{\mathrm{cat}}=1\, \mathrm{arcsec}$,
respectively. The analysis of the usual 50 Monte-Carlo runs resulted in
$\sigma_\gamma=2\cdot 10^{-3}$. We then scaled this value to the duration of the
Hipparcos mission, 3~yr, and to the number of observations expected for 44000 stars
(from the empirical law in Fig.1); this yielded $\sigma_\gamma=1\cdot 10^{-3}$.

That our experiment resulted in a much better value of $\sigma_{\gamma}$ should not come as a
surprise, simply because we put ourselves in a more favorable situation: the instrument and
satellite attitude were both assumed perfect in the simulation. In particular, having assumed a
perfect astrometric instrument (optics, focal plane, and detectors) we disregarded any possible
unmodeled systematic effect that could bias the estimation of $\gamma$, e.g., effects which would
mimic a parallax zero-point error, as it has been the case for Hipparcos (Lindegren et al. ~\cite{lennart},
Froeschl\'{e} et al.~\cite{froeschle}). In fact, in the observation equation derived from the
reduction model used in the Hipparcos $\gamma$ experiment, one can see that the unknown which represents
the parallax zero-point (common to all stars) is strongly correlated with the $\gamma$ parameter.
It can be shown that, for the Hipparcos mission parameters, such correlation amounts
to $\rho \simeq -0.92$ (Mignard~\cite{mignard}). A correlation of this magnitude, in turn, increases
the error in the estimate of $\gamma$ by a factor of $1/\sqrt{1 - \rho^2} ~\simeq 2.6$ ${}^($\footnote{While it is
extremely important that the astrometric parameters be free of this kind of systematic effects
which, though very small, could spoil any astrophysical result statistically inferred from such data,
the problem of $\gamma$ is of completely different nature. In other words, we should be
aware of systematic effects which correlate with the $\gamma$ parameter, but such effects need not
necessarily be modeled, provided they are smaller than the deviation of $\gamma$ from the GR value
we are trying to detect.}${}^)$. Therefore, the simulated ``replica" of the
Hipparcos experiment is to be considered consistent within a factor of 1.5,
and not a factor of 4, with the published results based on the real data; quite an encouraging
agreement given the quasi-ideal assumptions of our simulation.

\section{Summary and conclusions}

The global astrometric observations of a GAIA-like satellite were modeled within the PPN formulation
of Post-Newtonian gravitation. Although simplified (a spherical and non-rotating Sun is the
only source of gravity), this PPN model has allowed, through extensive end-to-end simulations,
a realistic evaluation of GAIA's sensitivity to the direct estimation of the light deflection
parameter. The results show that the satellite could measure $\gamma$ to $\sim 10^{-7}$
($1\sigma$) after 5 years of continuous observations, and using a subset of approximately
$10^6$ stars chosen as the most astrometrically stable among the millions available in the
magnitude range $V=12-13$ of the GAIA survey. Notice that after just one full year of observations
$\gamma$ could be estimated with an error only a factor of $\sqrt{2}$ worse than the value above.

A comparison with the Hipparcos results has provided a way to gauge the degradation factor to be expected
in going from an ideal instrument to the real astrometric payload and satellite. The factor of $1.5$ we found
in that case is very encouraging; however, moving from the ``milli-" to the ``micro-"arcsec regime
required by GAIA, the degradation might become larger. Future work will have to deal with the fact that
the accuracy of $10^{-7}$ sets a goal for both the observational model, which will have to include
all the details and the implications of Solar System gravitation, and instrument development and
modeling, which will have to concentrate in identifying and, possibly, remove (through hardware
improvements and/or calibration procedures), all relevant systematic error sources.

This work has provided quantitative evidence that the micro-arcsecond global astrometry of GAIA appears
capable of testing general relativity to unprecedented levels. It will do so directly by accurately
measuring the amount of light bending produced by gravity; a modern rendition, about a century
later, of the experiment of Dyson, Eddington, and Davidson, then the first proof of Einstein's theory.

\section{Acknowledgements}
We thank the referee for the valuable comments and for pointing us to
the more recent literature.

Work partially supported by the Italian Space Agency (ASI) under
contracts ASI I/R/32/00 and ASI I/R/117/01, and by the Italian
Ministry for Research (MIUR) through the COFIN 2001 program.


\end{document}